\documentstyle[prb,aps,preprint]{revtex}
\begin{document}
\draft
\title
{
Weak ferromagnetism of quasi-one-dimensional $S$ = 1/2 antiferromagnet 
BaCu${}_2$Ge${}_2$O${}_7$
}

\author{I. Tsukada,${}^1$ J. Takeya,${}^1$ T. Masuda,${}^{2}$ 
and K. Uchinokura${}^{2}$}
\address{${}^1$Central Research Institute of Electric Power Industry, 
2-11-1 Iwadokita, Komae-shi, Tokyo 201-8511, Japan}
\address{${}^3$Department of Advanced Materials Science, 
The University of Tokyo, 7-3-1 Hongo, Bunkyo-ku, Tokyo 113-8656, Japan}

\date{\today}
\maketitle

\begin{abstract}
Weak ferromagnetism of quasi-one-dimensional $S$ = 1/2 antiferromagnet 
BaCu${}_2$Ge${}_2$O${}_7$ is studied by the magnetization measurement.
The spontaneous magnetization appears along the $b$ axis below 8.8~K. 
The local symmetry between the intra-chain nearest neighbor spins allows 
the presence of Dzyaloshinskii-Moriya interaction, 
and the only possible spatial configuration of the weak ferromagnetic 
moment per spin uniquely determines the sign of the inter-chain interaction. 
A weak $a$-axis magnetic field can change the direction of the magnetization 
to the $a$-axis direction, which shows that the spin chain forms 
a weakly coupled weak-ferromagnetic chain system.

\end{abstract}

\narrowtext
One-dimensional (1D) quantum antiferromagnet (AF) is of 
particular interest in both theory and experiment of quantum magnetism. 
It is widely accepted that a pure 1D S=1/2 antiferromagnet has 
no long-range order at $T$ = 0~K.
\cite{Bethe1} 
On the other hand, almost all the actual spin-chain systems show 
magnetic three-dimensional long-range order (3D-LRO) at their ground states 
due to a finite inter-chain interaction. 
While in general there are many kinds of magnetic 3D-LRO states, 
the transition to a normal 3D AF-LRO state has been studied mostly. 
\cite{Hutchings1} 
One remarkable exception is a weak ferromagnetic transition that was 
recently found in 1D compound Sr${}_{0.73}$CuO${}_2$. 
\cite{Shengelaya1} 
However, this compound cannot be treated as a uniform spin-chain system 
but rather is well explained as an alternating-chain system. 
In this paper, we report weak ferromagnetism (WF) 
of a uniform $S$ = 1/2 spin chain in BaCu${}_2$Ge${}_2$O${}_7$, 
which is isomorphous to antiferromagnet BaCu${}_2$Si${}_2$O${}_7$.
\cite{Tsukada1} 
In BaCu${}_2$Ge${}_2$O${}_7$, intra-chain Dzyaloshinskii-Moriya (DM) 
interaction and inter-chain Heisenberg interaction cooperatively add 
a spontaneous magnetization in contrast to AF BaCu${}_2$Si${}_2$O${}_7$.

The single-crystal sample of BaCu${}_2$Ge${}_2$O${}_7$ was grown by 
a floating-zone method, and was cut into a rectangular shape ($a$ $\times$ $b$ 
$\times$ $c$ = 1.7 $\times$ 1.6 $\times$ 3.1 mm${}^3$). 
Since the $a$- and the $c$-axis lengths are almost the same, 
particular attention was paid to determine the crystal orientation by 
x-ray diffraction. We took the diffraction at randomly selected three 
different points at every surface, and found no trace of a twinned 
structure. However, the possibility of inclusion of misoriented 
grains inside the crystal was not eliminated. 
Magnetization measurement was carried out with a commercial SQUID 
magnetometer (MPMS, Quantum Design) in the field range from 0 to 5~T. 
We also measured transverse magnetization with this system.

Figure~\ref{Fig.1} shows the magnetic susceptibility 
measured at $H$ = 1000~Oe. 
A broad peak is observed around 300~K, and the data can be fitted 
with a calculation by Bonner and Fisher (BF),
\cite{Bonner1} 
which indicates that the system is well described as an 1D Heisenberg AF. 
$\chi_b$ and $\chi_c$ are well fitted assuming different $g$ values, and 
$J$ = 540~K is obtained as an intra-chain coupling showing a good agreement 
with that for the polycrystalline sample.
\cite{Tsukada1} 
As the temperature decreases to the WF transition temperature 
at $T_N$ = 8.8~K, 
deviation from the BF curves becomes significant. Below $T_N$, 
spontaneous magnetization appears as shown in the inset of Fig.~\ref{Fig.1}. 
The magnetization along the $b$ axis ($M_b$) shows 
a typical evolution for ferromagnetic transition 
and is saturated at approximately 9$\times$10${}^{-2}$ emu/g 
for $T$ $\rightarrow$ 0, which corresponds only to 0.42{\%} of 
what we may observe when all the $S$ = 1/2 spins on 
Cu${}^{2+}$ site are aligned toward the same direction. 
Thus we eliminate the possibility of normal 
ferromagnetic transition. Ferrimagnetism is impossible, 
because BaCu${}_2$Ge${}_2$O${}_7$ has no magnetic ions other 
than Cu${}^{2+}$ that occupies an equivalent site. 
Since the sample is highly insulating, itinerant or 
band ferromagnetism is impossible.

To see the anisotropic magnetization, the field dependence 
of magnetization was measured at the WF state. 
Figure~2 shows low-field magnetization along three principal axes. 
A characteristic steep increase of the magnetization is found 
along the $a$ and $c$ axes up to $H$ = 280~Oe ($H_c$). 
$M_a$ reaches more than 1.6 $\times$ 10${}^{-1}$ emu/g, 
while $M_b$ remains approximately 9$\times$10${}^{-2}$ emu/g. 
Once magnetic field exceeds $H_c$, the slopes become similar to one another 
as shown in the inset of Fig.~\ref{Fig.2}. 
The magnetization continues to increase along all the axes, 
which is a typical behavior of WF-LRO state. 
We did not observe a spin-flop anomaly in this field region, 
which indicates the lack of a strong easy-axis anisotropy in this system.

Let us first discuss the crystal structure to reveal this WF-LRO state. 
BaCu${}_2$Ge${}_2$O${}_7$ has a structure isomorphous to 
BaCu${}_2$Si${}_2$O${}_7$ (see Fig.~1 of Ref.~4) with slightly longer 
lattice parameters: $a$ = 7.028~{\AA}, $b$ = 13.403~{\AA}, 
and $c$ = 7.044~{\AA}.
\cite{Oliveira1}
Two neighboring Cu${}^{2+}$ ions are connected by one O${}^{2-}$ ion 
and they form a chain along the $c$ axis. However, the bond of Cu-O-Cu 
is not straight as shown in Fig.~\ref{Fig.3} (a). 
The angle of $\angle$ Cu-O-Cu (135${}^{\circ}$) is larger than that 
of BaCu${}_2$Si${}_2$O${}_7$ (124${}^{\circ}$), which is consistent with 
the larger intra-chain interaction. 
Two mechanisms of WF are known up to now: 
one is a single-ion anisotropy for NiF${}_2$,
\cite{Haendler1} 
and the other is a DM antisymmetric interaction for $\alpha$-Fe${}_2$O${}_3$,
\cite{Neel1} 
where the antisymmetric exchange interaction 
${\bf D}\cdot[{\bf S}_i{\times}{\bf S}_{i+1}]$ gives spin canting.
\cite{Dzyaloshinskii1} 
There cannot be a single-ion anisotropy in BaCu${}_2$Ge${}_2$O${}_7$, 
because a magnetic moment in BaCu${}_2$Ge${}_2$O${}_7$ is attributed 
to a $S$ = 1/2 spin on Cu${}^{2+}$ ions. 
Then it necessarily follows that the DM interaction is the 
origin of WF in this compound. 
The local symmetry of nearest neighbor (nn) Cu${}^{2+}$ ions is 
actually low enough to allow the DM interaction. 
According to the rule given by Moriya,
\cite{Moriya1}
the Dzyaloshinskii vectors ${\bf D}_i$ at bond $i$ are approximately 
defined as schematically drawn in Fig.~\ref{Fig.3} (a). 
The positions of only two neighboring Cu${}^{2+}$ ions and one 
intermediate O${}^{2-}$ ion are taken into account. 
According to the room-temperature structural data,
\cite{Oliveira1} 
${\bf D}_i$ is written as 
${\bf D}_i = D$ ((-1)${}^i$0.8332, (-1)${}^i$0.5462, 0.0860). 
Here, the components perpendicular and parallel to the chain have 
different effects on the spin order; 
the former (${\bf D}_i^{\perp}$ = (-1)${}^iD$(0.8332, 0.5462, 0)) changes 
its sign from one bond to the next, while the latter 
(${\bf D}_i^{\parallel}$ = $D$(0, 0, 0.0860)) keeps its sign along the chain. 
For the rest of the paper we shall ignore the effect of 
${\bf D}_i^{\parallel}$ and simply consider only ${\bf D}_i^{\perp}$ from 
the following two reasons: 
1) the magnitude of ${\bf D}_i^{\parallel}$ is far smaller than that of 
${\bf D}_i^{\perp}$, and 
2) ${\bf D}_i^{\parallel}$ will stabilize a spiral order that is 
against the experimental observation.

We must know the arrangement of D vectors in the whole magnetic 
unit cell, because the bulk spontaneous magnetization is closely 
related to how D vectors are distributed. It is noted that 
D vectors and spin operators are both an axial vector 
when we consider the symmetry operations.
For the $a$-axis directions, one chain is transformed to the neighboring 
one by the successive operations of a) reflection in the ac plane and 
b) rotation around the axis located at the middle of two adjacent Cu${}^{2+}$ 
ions along the [101] direction and parallel to the $b$ axis. 
Through these operations, the $a$ component of ${\bf D}_i^{\perp}$ changes 
its sign twice, the $b$ component is unchanged, and the order of the 
concerning spin operators are exchanged. As a result, ${\bf D}_i^{\perp}$ 
at the $a$-axis neighboring bonds points an opposite direction. 
We can also transform a chain to the $b$-axis neighboring one 
by the operation of reflection in the $ac$ plane and following parallel shift 
to the $b$ direction. This process changes the sign of the $a$ component only. 
In this way we determined all the D vectors in the magnetic unit cell 
as shown in Fig.~\ref{Fig.3} (b).

Next we discuss how these D vectors tilt spins to create the $b$-axis 
spontaneous magnetization. For this purpose, we must determine the principal 
direction of the spins. The crystal symmetry is high enough that it must be 
along the $a$, $b$, or $c$ direction. 
If we take only DM interaction into account, the fact that D vectors are not 
collinear along the $b$ direction can differentiate the three cases because 
there is a finite inter-chain interaction along the $b$ axis. 
However, the recent theoretical studies proved that DM interaction is always 
accompanied by an additional term quadratic to $D$, which recovers 
$O(3)$ symmetry of the spin-spin interaction.
\cite{Kaplan1}
This term is called as KSEA interaction, and has been established also 
experimentally.
\cite{Zheludev1}
Therefore we cannot predict the principal spin direction from DM and KSEA 
interactions even in the presence of inter-chain interaction.

On the other hand, experimental results indicates that the principal spin 
direction is along the $c$ axis. 
This is explained as follows; 
1) Each chain must have weak ferromagnetic moment along the $b$ direction, 
while ${\bf D}_i^{\perp}$ has the $a$ and $b$ components, 
2) if the principal direction is along the $a$ or $b$ direction, spins are 
canted only toward the $c$ direction and we will have no $b$-axis 
magnetization, 
3) only when the principal direction is along the $c$ axis, we will observe 
weak ferromagnetic moment along both the $a$ and $b$ directions. 
Thus, we safely concluded that the $c$ axis is the easy-axis direction 
in the actual system. 
Under this condition, spins on a single chain are confined in the plane 
perpendicular to D vectors (we shall call it as {\em easy plane} for 
convenience) as shown in Fig.~\ref{Fig.3} (c). 
It is convenient to define the WF moment per chain as the sum of 
the two sub-lattice magnetizations: 
${\bf M}_{WF,k}$ = (${\bf M}_{1,k} + {\bf M}_{2,k}$)/2, 
where ${\bf M}_{i,k}$ denotes the magnetic moment at the $i$-th 
sub-lattice in the $k$-th chain. 
Obviously the direction of the weak ferromagnetic moment 
is parallel to the easy-plane and simultaneously perpendicular to the $c$ 
axis. Once we obtain the direction of ${\bf M}_{WF,k}$ on each spin chain, 
the spatial configuration of ${\bf M}_{WF,k}$ determines the bulk 
magnetization. Relative directions of ${\bf M}_{WF,k}$ between the neighboring 
chain depend on the inter-chain interaction. 
In Fig.~\ref{Fig.3} (d), four different cases according to the signs 
of $J_a$ and $J_b$ are schematically drawn. 
Among the four cases, a spontaneous magnetization along the $b$ axis 
is realized only when $J_a >$ 0 and $J_b >$ 0, 
and therefore BaCu${}_2$Ge${}_2$O${}_7$ is considered to be the case. 
Consequently, the spin configurations in the magnetic unit cell os 
determined as shown in Fig.~\ref{Fig.3} (e).

So far, we have assumed that the DM interaction is the dominant perturbation 
to the spin system and the inter-chain interaction gives only a secondary 
effect. This assumption is probably valid by the following discussion. 
The magnitude of DM interaction can be estimated from the magnetization data. 
Using the angle between the easy plane and the $ac$ plane being 
57.3${}^{\circ}$, the canting angle of each spin from the 
$c$-axis direction is estimated as $\theta$ = 0.95${}^{\circ}$.
\cite{mu1} 
Then the coarse relation, $\tan{2\theta} \cong D / J$, gives $D$ as 18~K. 
On the other hand, we can also estimate the inter-chain interaction 
as an averaged value according to the chain mean-field theory:
\cite{Schulz1} 
$J$ = 540~K and $T_N$ = 8.8~K give the averaged inter-chain 
interaction as ${\mid}J^{\perp}{\mid}$ = 2.9~K, 
which is sufficiently smaller than $D$. 
Of course, the direct estimation of $J_a$ and $J_b$ is necessary for 
further quantitative discussion, but $|D| {\gg} |J_a|$ and 
$|D| {\gg} |J_b|$ are probably valid in BaCu${}_2$Ge${}_2$O${}_7$. 
The above discussion means that BaCu${}_2$Ge${}_2$O${}_7$ is a good system 
to study the magnetism of weak-ferromagnetic spin chains that are 
weakly coupled with each other, 
and the sign of the inter-chain interaction can easily change the 
magnetic ground state demonstrated as the difference between 
BaCu${}_2$Ge${}_2$O${}_7$ and BaCu${}_2$Si${}_2$O${}_7$.

Finally we discuss the low-field anisotropic magnetization of 
BaCu${}_2$Ge${}_2$O${}_7$ upon this model structure. 
$M_a$ steeply increases up to $H_c$ first and suddenly changes its slope. 
This is qualitatively explained by the rotation of WF moment 
as schematically shown in Fig.~\ref{Fig.4} (a). 
When the field is applied parallel to the $a$ axis, 
${\bf M}_{WF,1}$ and ${\bf M}_{WF,2}$ hold their directions, 
while ${\bf M}_{WF,3}$ and ${\bf M}_{WF,4}$ are rotated almost by 
180${}^{\circ}$, which can produce a large magnetization along the $a$ axis. 
We consider that the spins on the chains 3 and 4 are cooperatively 
rotated without changing their relative angle so much, 
because both $J_b$ and the Zeeman energy are too 
weak to change the relative angle of the spins against the 
dominant DM interaction. 
After this rotation has been finished at 280~Oe, large magnetization appears 
along the $a$ axis, and {\em simultaneously} net magnetization along 
the $b$ axis should disappear. In order to confirm this rotation, 
we carried out the transverse magnetization measurement, and found 
that the $b$-axis magnetization is actually suppressed with increasing 
field along the $a$ axis as shown in Fig.~\ref{Fig.4} (b). 
Since the geometry of the pick-up coil for our transverse magnetization 
measurement was the second differential type, 
finite $M_b$ gives an even-function response to the direct signal. 
Actually we got such a signal around 0~Oe, and observed the evolution 
of $M_b$. However, as the field approach 280~Oe, the direct signal changes 
its shape and finally turns to an odd-function response, which means that the 
$M_a$ induced by the field dominates the response of the pick-up coil. 
$M_b$ is now negligibly small, and we conclude that the spin structure 
shown in Fig.~\ref{Fig.4} (b) is realized. 
Such spin rotation is one of the characteristics of weakly-coupled 
spin chain, where DM interaction roughly keeps WF moment per chain, 
while the direction of WF moment is determined as a result of 
the competition between Zeeman energy and inter-chain coupling.

The difference of the magnetic ground states in BaCu${}_2$Ge${}_2$O${}_7$ and 
BaCu${}_2$Si${}_2$O${}_7$ provides us useful spin system to study the effect 
of randomness in the inter-chain interaction. 
Both compounds are the end materials of the solid-solution system 
of BaCu${}_2$(Ge${}_{1-x}$Si${}_x$)${}_2$O${}_7$. 
The mixture of Ge and Si will introduce randomness to the sign of the 
inter-chain coupling along the $a$ axis. This randomness reduces the evolution 
of 3D long-range order as is reported in a polycrystalline samples 
by Yamada and Hiroi.
\cite{Yamada1}

To summarize, we discovered the weak-ferromagnetic state in the 
1D Heisenberg antiferromagnet BaCu${}_2$Ge${}_2$O${}_7$, 
which is to our knowledge the first uniform $S$ = 1/2 spin chain system 
that shows weak-ferromagnetic long-rang order. 
The Dzyaloshinskii-Moriya interaction adds a weak-ferromagnetic moment 
to each spin chain, and the relatively weak inter-chain interaction 
allows us to treat this system as weakly-coupled weak-ferromagnetic chains. 
The spin rotation by weak external field becomes possible by the combination 
of weak ferromagnetic moment per each spin 
and relatively weak inter-chain interaction, which is 
expected only in such a weakly coupled 1D spin system.

We wish to acknowledge valuable discussions with A. Zheludev, T. Yamada, 
and Z. Hiroi. I.T. would like to thank also Y. Ando for renting me a SQUID 
magnetometer. Work at the University of Tokyo is supported in part 
by the Grant-in-Aid for COE Research by the Ministry of Education, Science, 
Sports, and Culture of Japan.

\begin{figure}[t]
\caption
{Magnetic susceptibilities along the three principal axes 
measured at 1000 Oe. The solid and dashed lines show the calculation 
by Bonner and Fisher with $J$ = 540~K as a fitting parameter assuming a 
different $g$ values. 
Inset: Low-temperature magnetization along the three principal axes 
measured at $H$ = 10~Oe. 
}
\label{Fig.1}
\end{figure}

\begin{figure}
\caption
{Field dependence of the magnetization along the principal axes below 
1000~Oe at $T$ = 5~K. 
At $H$ = 0~Oe, spontaneous magnetization appears along the $b$ axis, 
while the magnetization along the $a$ axis becomes the largest above 180~Oe 
and has a kink at 280~Oe. 
Inset: Field dependence of the magnetization up to 5~T 
at the same temperature. 
}
\label{Fig.2}
\end{figure}

\begin{figure}
\caption
{(a) Single Cu-O chain, and {\bf D} vectors between the nn Cu${}^{2+}$ ions 
deduced from the local symmetry . 
(b) 3D schematic arrangement of ${\bf D}_{\perp}$'s in the magnetic 
unit cell. All the vector products are calculated from left to right as 
indicated by the arrow. See details in the text. 
(c) Model of spin canting on the easy plane. The WF moment per spin chain 
is defined as ${\bf M}_{WF,k} = ({\bf M}_{1,k} + {\bf M}_{2,k}) / 2$. 
(d) Spatial configuration of ${\bf M}_{WF,k}$ for [$J_a >$ 0, $J_b >$ 0], 
[$J_a <$ 0, $J_b >$ 0], [$J_a >$ 0, $J_b <$ 0], and [$J_a <$ 0, $J_b <$ 0]. 
(e) 3D spin arrangements of BaCu${}_2$Ge${}_2$O${}_7$. 
}
\label{Fig.3}
\end{figure}

\begin{figure}
\caption
{(a) Schematic picture of spin rotation by the $a$-axis field. 
${\bf M}_{WF,k}$ is indicated by an arrow. 
(b) Transverse magnetization measurement. Magnetic field is applied along 
the $a$ axis and the magnetization along the $b$ axis is measured. 
The inset shows a direct response of the pick-up coil.
}
\label{Fig.4}
\end{figure}

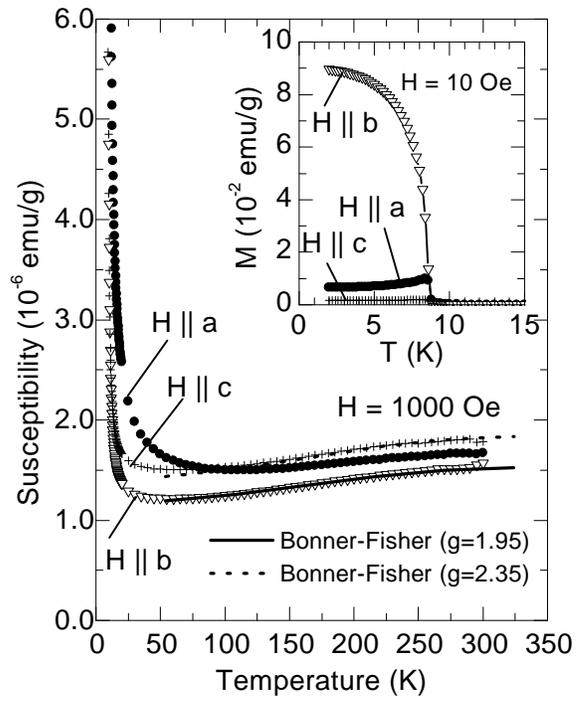
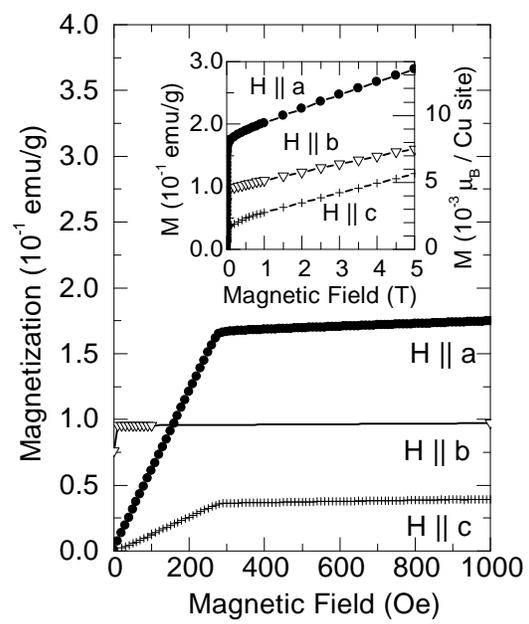
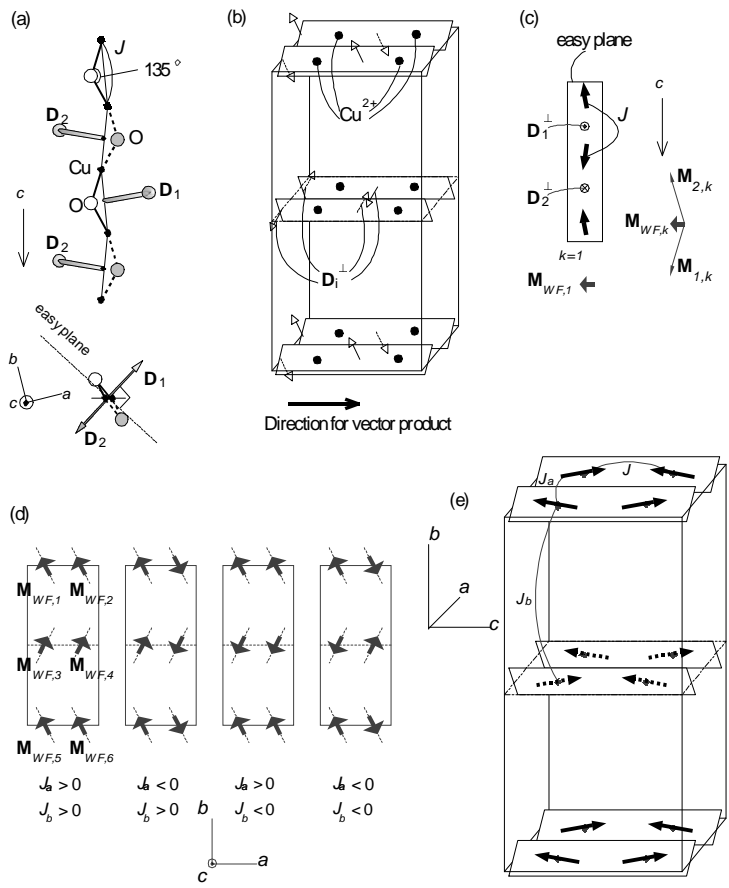
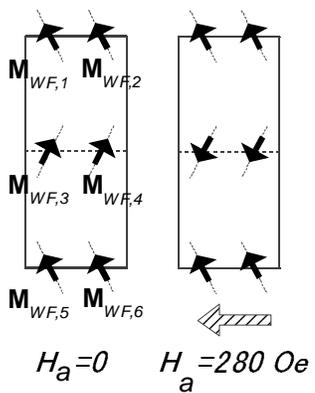
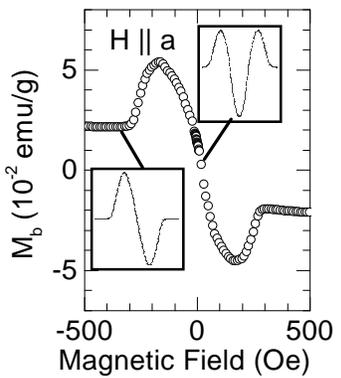
\end{document}